\documentclass[traditabstract]{aa}  

\usepackage{graphicx}
\usepackage{txfonts}
\usepackage{epsfig}
\usepackage{natbib}

\begin{document}

\title{H-alpha monitoring of OJ~287 in 2005-08 \thanks{Based on
observations collected at the European Organization for Astronomical
Research in the Southern Hemisphere, Chile, proposals
075.B-0527, 076.B-0028, 077.B-0015, 078.B-0030, 079.B-0014, 380.B-0665 and
381.B-0050.}}

\author{
K. Nilsson\inst{1}\fnmsep\inst{2}\and
L. O. Takalo\inst{1}\and
H. J. Lehto\inst{1}\and
A. Sillanp\"a\"a\inst{1}
}
   
\institute{Tuorla Observatory, Department of Physics and Astronomy,
University of Turku, V\"ais\"al\"antie 20, FI-21500 Piikki\"o, Finland
\and
Finnish Centre for Astronomy with ESO (FINCA), University of Turku,    
V\"ais\"al\"antie 20, FI-21500 Piikki\"o, Finland
}

\date{Received; accepted}

\abstract { 

We present the results of H$\alpha$ monitoring of the BL Lac object
OJ~287 with the VLT during seven epochs in 2005-08. We were able to detect
five previously undetected narrow emission lines,
$\lambda\lambda$6548, 6583[NII], $\lambda6563$H$\alpha$ and
$\lambda\lambda$6716, 6731 [SII] during at least one of the epochs and
a broad H$\alpha$ feature during two epochs.  The broad H$\alpha$
luminosity was a factor $\sim$ 10 lower in 2005-08 than in 1984 when
the line was previously detected and a factor $\sim$ 10 lower than
what is observed in quasars and Seyfert galaxies at the same
redshift. The data are consistent with no change in the position
or luminostity of the H$\alpha$ line in 2005-08.
The width of the H$\alpha$ line was $4200 \pm 500$ km
s$^{-1}$, consistent with the width in 1984. 
}


\maketitle

\section{Introduction}

The BL Lac object \object{OJ 287} (z = 0.306) has received some
attention after the discovery of the recurring outbursts every $\sim$
12 years in the optical data over the past century
\citep{1988ApJ...325..628S}.  This observation has led many authors to
consider a binary black hole model for OJ~287
\citep{1988ApJ...325..628S,1996ApJ...460..207L,1997ApJ...478..527K,1997ApJ...484..180S,1998MNRAS.293L..13V,2000ApJ...531..744V,2006ApJ...646...36V,2007ApJ...659.1074V}.
The exact mechanism of the outburst differs from one model to another.
Some models attribute the flux increase to changes of the direction of
the relativistic jet \citep[e.g.][]{1997ApJ...478..527K}, other models
to the impact of the secondary black hole onto the accretion disk of
the primary followed by tidal effects and increased luminosity of the
jet \citep[e.g.][]{1996ApJ...460..207L}. The most detailed model so
far \citep{2007ApJ...659.1074V} suggests a very high mass for the
primary black hole $M = 1.8 \times 10^{10}\ M_{\odot}$ and a very
eccentric orbit for the secondary ($e = 0.66$). To explain the
aperiodicity of the outbursts the precession of the orbit has to be
very strong, $\Delta \phi = 40^{\circ}$ per orbit, in this model.

Testing these models was so far concentrated mainly on the timing of
the outbursts with broadband photometric observations
\citep[e.g.][]{2009ApJ...698..781V} with other aspects receiving less
attention. One interesting constraint to the models could be obtained
by monitoring the emission line variability during the outbursts. The
luminosity of the emission lines could change in response to varying
continuum radiation or, if the size of the secondary black hole orbit
and the characteristic broad line region (BLR) size are of the same
order, the secondary black hole could affect the velocity field of the
BLR.  Furthermore, if the BLR velocity dispersion was measured, it
could be used to constrain the mass of the primary black hole in
OJ~287.  However, there are no published results of spectroscopic
monitoring of OJ~287 in the literature.  Thus we started a project in
2005 at the VLT to monitor the broad line region of OJ~287 during the
expected outbursts in 2005-08 when the two black holes were expected
to be at closest approach by various models.  The main aim of this
project was to see if any changes could be observed in the BLR of
OJ~287 during the expected closest approach of the binary.

This kind of monitoring project is quite challenging because the
emission lines are very weak in BL Lacs by definition.  There are only
two reported detections of broad lines in OJ~287 in the
literature. \cite{1989A&AS...80..103S} detected the H$\beta$ line
with a FWHM of 56 \AA\ with an equivalent width (EW) of 1.1
\AA. \cite{1985PASP...97.1158S} also detected a weak H$\beta$, but
the most notable feature in their spectrum is the prominent broad
(FWHM $\sim$ 120 \AA) H$\alpha$ line. It is impossible to know if
these detections represent typical levels or very high states of the
lines. Given the weakness of emission lines in BL Lacs in general, the
latter alternative seems more probable as non-detections are not
always reported. Based on this past information the H$\alpha$ line
was chosen for the monitoring because it can be easily detected with
modern spectrographs at sufficiently large telescopes if the
luminosity of the line is close to what was observed by
\cite{1985PASP...97.1158S}. The results of this monitoring are
reported in this paper.

Throughout this paper we use the cosmology $H_0 = 70$ km s$^{-1}$
Mpc$^{-1}$, $\Omega_{M}$ = 0.3 and $\Omega_{\Lambda}$ = 0.7.

\section{Observations and data reduction}

The observations were made with the FORS2 instrument
\citep{1998Msngr..94....1A} attached to the VLT-UT1 (Antu) in service
mode during seven epochs in 2005-08. Throughout the whole period the same
instrumental setup was used, employing the 1028z grism with the OG590
order separation filter. The wavelength range covered by the 1028z
grism is $\lambda\lambda$ 7730-9480 \AA\ and the resolution with the
2\arcsec slit was 8.1 \AA.  The observations are summarized in Table
\ref{obstable}. The exposure time in Table \ref{obstable} gives the
total exposure time of 2-5 exposures of OJ~287 made a each epoch.

The reduction of the spectra was made with IRAF.  First a bias frame
was subtracted from the two-dimensional CCD spectra after which they
were divided by a flat field frame derived from continuum lamp
exposures. A two-dimensional wavelength calibration solution was
derived from strips of data separated by 5 arcsec in the spatial
direction using the same 14 night-sky emission lines in each
strip. This solution was used to rectify the image before subtracting
the night-sky emission lines. After the night-sky subtraction the
one-dimensional spectrum of OJ~287 was extracted by tracing the
spectrum of OJ~287 on the CCD and integrating the light within a 7.5
arcsec aperture. Optimal extraction was used at this phase to
obtain maximum signal-to-noise.

The spectra were calibrated using five spectrophotometric calibration
stars (LTT 4816, EG 274, LTT 7379, LTT 6248 and GD 108) obtained with
the same instrumental setup as here during the observations or from
the ESO archive observations made in 2005-06.  The average sensitivity
curve was derived form these five stars and applied to the spectra of
OJ~287. An extinction correction was made to each spectrum using a
standard atmospheric extinction curve and the airmass at the time of
the observation. Some of the nights were affected by thin clouds,
whose absorption cannot be accounted for by the above calibration
procedure.

The calibrated spectra exhibit absorption bands from atmospheric water
vapor, which are particularly prominent at $\lambda\lambda$8100-8400
\AA\ and $\lambda\lambda$8800-9300 \AA\ and whose strength varied from
one epoch to another. To correct for the atmospheric absorption bands
a template was derived from the spectrum of the nearby star 10
\citep{1996A&AS..116..403F} observed simultaneously with OJ~287 on
2008-01-05. This template was scaled appropriately to obtain the best
correction for each epoch.  During this phase it became evident that
due to the variable nature of the atmospheric absorption bands they
cannot be completely removed at their cores. The wavelength regions
most affected by this effect were excluded from further analysis (see
below).

\begin{table}
\caption{\label{obstable}Summary of the observations. The last
column gives the signal-to-noise at 8700 \AA.}
\centering
\begin{tabular}{llllll}
\hline
Epoch & Date & t$_{\sf exp}$ & FWHM & Airmass & S/N\\
 & & (s) & (arcsec) & &\\
\hline
1 & 2005-04-05 & 1200 & 0.9 & 1.46 & 280\\
2 & 2005-11-19 & 1800 & 0.9 & 1.61 & 330\\
3 & 2006-04-02 & 1500 & 0.6 & 1.41 & 400\\
4 & 2006-12-06 & 1800 & 0.9 & 1.50 & 315\\
5 & 2007-04-04 & 2500 & 0.7 & 1.43 & 440\\
6 & 2008-01-05 & 2500 & 0.8 & 1.41 & 560\\
7 & 2008-04-08 & 2640 & 1.1 & 1.42 & 410\\
\hline
\end{tabular}
\end{table}

\section{Analysis and results}

\begin{figure}
\centering
\epsfig{file=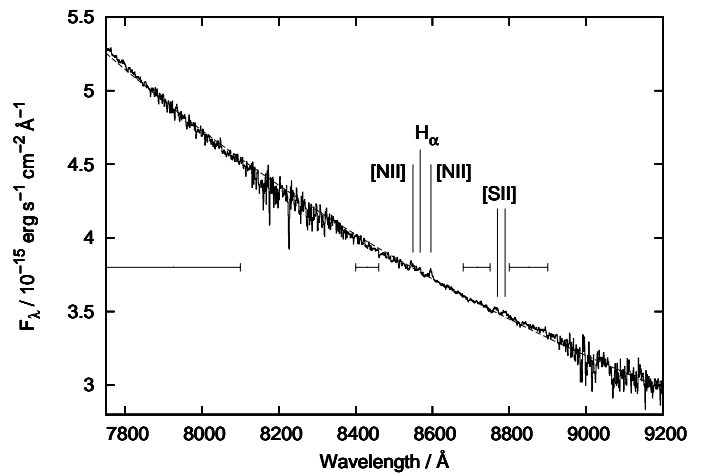,width=9cm}
\caption{\label{spectrum} Spectrum of OJ~287 on 2008-01-05. The
dashed line is a power-law fit to pure continuum parts of
the spectrum, indicated by horizontal bars. The spectral lines
identifiable in the spectrum have also been marked.}
\end{figure}

Figure \ref{spectrum} shows an example of a calibrated spectrum of
OJ~287 obtained during the monitoring campaign. The spectrum shows two
noisy areas at $\lambda\lambda$8100-8400 \AA\ and
$\lambda\lambda$8850-9200 \AA\ due to the imperfect subtraction of the
atmospheric absorption bands. The spectra also show two areas
populated by weak emission lines: at $\lambda \sim$ 8570 \AA\ is the
$\lambda\lambda$6548,6583 [NII] + $\lambda6563$ H$\alpha$ group and
at $\lambda \sim$ 8780 \AA\ the $\lambda\lambda$6716,6731 [SII] lines.
Horizontal bars in Fig. \ref{spectrum} indicate the wavelength areas
representing ``pure'' continuum. These areas, 7750-8100 \AA, 8400-8460
\AA, 8680-8750 \AA\ and 8800-8900 \AA\ are used in all subsequent work
for continuum fitting.

\begin{figure}
\centering
\epsfig{file=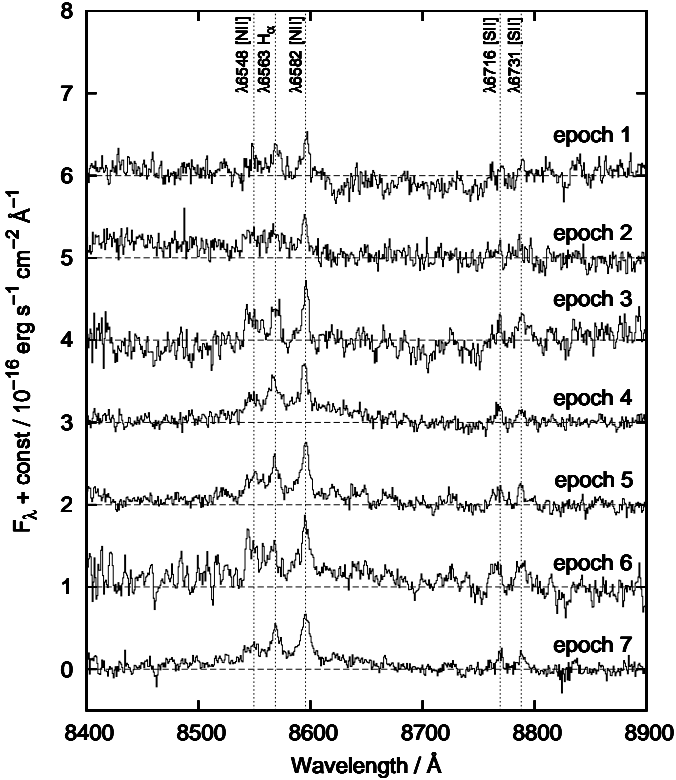,width=9cm}
\caption{\label{spektrit} Continuum-subtracted spectra of OJ~287
during the monitoring. Individual spectra have been shifted by $10^{-16}$
erg s$^{-1}$ cm$^{-2}$ \AA$^{-1}$ in vertical direction for clarity.}
\end{figure}

The overall shape of the continuum follows a power-law very well as
indicated by the power-law fit in Fig. \ref{spectrum}. In all seven
spectra the maximum deviation from the power-law fit is 1\%. However,
because the main interest of this work is on the line variations of
OJ~287, a second order polynomial was chosen for the continuum fit as
it gives about two times smaller residuals than the power-law fit and
still retains the smooth and monotonous shape of the power-law
spectrum.  Figure \ref{spektrit} shows the continuum-subtracted
spectra during the seven epochs of the monitoring. We note that the
continuum level varied by a factor of 2.6 during the monitoring.

During at least two of the epochs (four and seven) a broad spectral
feature is seen at the location of the H$\alpha$-[NII] lines and
epochs five and six show a hint of a similar feature. However, as is
evident from Fig. \ref{spektrit}, the continuum level is fluctuating
around the mean level due to calibration errors etc. Because these
fluctuations could be wrongly interpreted as true features, it is
important to study the statistical properties of the fluctuations more
closely.

To study the significance of the spectral features in
Fig. \ref{spektrit} we employed the method by
\cite{2005AJ....129..559S}.  This method consists of dividing the
spectrum into spectral bins, computing the equivalent width (EW) of
each bin and deriving the rms fluctuations of the EW through the
spectrum. Any EW exceeding a pre-defined rms limit (e.g. 3$\sigma$) is
taken as a sign of significant emission or absorption line warranting
further study.  \cite{2005AJ....129..559S} computed the continuum
level for each bin from two adjacent bins, which works fine for narrow
lines, but very broad features may be missed. Thus the method by
\cite{2005AJ....129..559S} was slightly modified by using the fitted
continuum value instead of the adjacent bins as the continuum
estimate.  The rms noise in the EW was computed from the pure
continuum areas marked with horizontal bars in Fig. \ref{spectrum}. A
bin size of 8 \AA\ was used in the computations and as in
\cite{2005AJ....129..559S} any EW value thrice the rms EW was
interpreted as a significant spectral feature. The two areas where the
atmospheric line correction produces too much noise
(Fig. \ref{spectrum}) were excluded from the analysis.

The results of this analysis are shown in Table \ref{merkitys}.
Column 3 of the table gives the derived 3$\sigma$ EW limit used to
detect significant features. Columns 3-7 list whether the five narrow
emission lines seen in the spectrum were formally detected or not. The
last two columns give wavelength regions where broad spectral
features, defined as at least three adjacent bins with significant
emission, were detected and the last column gives any other significant
spectral features detected.

\begin{table*}
\caption{\label{merkitys} Results of the search for significant
spectral features.}
\centering
\begin{tabular}{lllcccccll}
\hline
      &              &          & \multicolumn{5}{c}{Detected?} & Broad & Other\\
Epoch & JD - 2450000 & EW limit & [NII] & H$\alpha$ & [NII] & [SII] & [SII] & features &
features\\
      & & (\AA)              & 6548 & 6563 & 6583 & 6716 & 6731 & &\\
      & & (\AA\ redshifted) & 8550 & 8569 & 8595 & 8769 & 8789 & &\\
\hline
1 & 3466.56759 & 0.082 & no  & yes & yes & no  & no  & \ldots & absorption at 8738 \AA\\
2 & 3693.83049 & 0.092 & no  & no  & yes & no  & no  & \ldots & \ldots\\
3 & 3827.53441 & 0.072 & no  & no  & yes & no  & no  & \ldots & \ldots\\
4 & 4075.80737 & 0.059 & yes & yes & yes & yes & yes & emission at 8538-8642 \AA &\ldots\\
5 & 4194.55954 & 0.051 & yes & yes & yes & no  & yes & emission at 8546-8586 \AA & absorption at 8058\AA\\
6 & 4470.76465 & 0.067 & yes & yes & yes & no  & no  & emission at 8546-8570 \AA &\ldots\\
7 & 4564.53963 & 0.061 & yes & yes & yes & no  & no  & emission at 8538-8602 \AA &\ldots\\
\hline
\end{tabular}
\end{table*}

Of the five visually identified narrow emission lines only $\lambda$6583
[NII] is consistently detected throughout the campaign. The rest are
detected more sporadically, especially the [SII] lines are mostly
below the 3$\sigma$ level. There is a clear indication of broad
emission on top of the [NII]/H$\alpha$ lines, which is most
naturally explained as broad H$\alpha$ emission. This broad
emission component is particularly clear at epoch four. In addition to
the lines in Table \ref{merkitys} the only detected significant
features are two weak (3$\sigma$) absorption features at 8058 and 8738
\AA. These lines do not correspond to any commonly observed spectral
lines at the redshift of OJ~287.

\begin{table}
\caption{\label{voima} Results of the line fitting.}
\centering
\begin{tabular}{llccc}
\hline
Epoch & line & FWHM & $\lambda_c$ & Flux\\
      &      & (\AA)  & (\AA) & ($10^{-16}$ erg\\
      &      &      &     & s$^{-1}$ cm$^{-2}$)\\
\hline
1 & H$\alpha$ narrow & 6.1$^{+3.1}_{-2.0}$ & 8569.3$^{+1.0}_{-0.6}$ &  4.0$^{+2.0}_{-1.6}$\\
  & 6583 [NII]          &                     & 8596.1$^{+0.8}_{-1.1}$ &  4.4$^{+2.1}_{-1.7}$\\
2 & 6583 [NII]          & 7.7$^{+7.2}_{-3.3}$ & 8594.6$^{+1.2}_{-0.9}$ &  5.8$^{+5.8}_{-3.2}$\\
3 & 6583 [NII]          & 4.7$^{+2.7}_{-1.7}$ & 8595.5$^{+1.1}_{-0.7}$ &  5.3$^{+2.9}_{-2.0}$\\
4 & 6548 [NII]          & 6.4$^{+1.4}_{-1.2}$ & 8549.1$^{+2.0}_{-1.6}$ &  2.1$^{+0.9}_{-0.7}$\\
  & H$\alpha$ narrow &                     & 8567.9$^{+1.1}_{-1.1}$ &  3.8$^{+1.0}_{-1.0}$\\
  & H$\alpha$ broad  & 106$^{+22}_{-14}$   & 8589.7$^{+15.5}_{-12.4}$ & 22.5$^{+3.8}_{-4.1}$\\
  & 6583 [NII]          &                     & 8594.2$^{+0.9}_{-0.9}$ &  4.9$^{+1.3}_{-1.1}$\\
  & 6716 [SII]          &                     & 8767.0$^{+1.1}_{-1.4}$ &  1.7$^{+0.7}_{-0.6}$\\
  & 6731 [SII]          &                     & 8787.9$^{+1.3}_{-1.9}$ &  1.5$^{+0.7}_{-0.5}$\\
5 & 6548 [NII]          & 7.4$^{+2.0}_{-1.6}$ & 8549.8$^{+2.0}_{-1.6}$ &  3.4$^{+1.8}_{-1.3}$\\
  & H$\alpha$ narrow &                     & 8567.5$^{+1.7}_{-1.4}$ &  4.5$^{+1.8}_{-1.6}$\\
  & 6583 [NII]          &                     & 8594.2$^{+1.2}_{-1.0}$ &  6.7$^{+1.9}_{-1.8}$\\
  & 6731 [SII]          &                     & 8788.5$^{+1.8}_{-2.2}$ &  2.4$^{+1.1}_{-0.9}$\\
6 & 6548 [NII]          & 8.9$^{+9.9}_{-3.0}$ & 8547.1$^{+9.0}_{-3.2}$ &  5.7$^{+4.0}_{-2.9}$\\
  & H$\alpha$ narrow &                     & 8564.7$^{+5.8}_{-8.4}$ &  4.3$^{+3.8}_{-3.7}$\\
  & 6583 [NII]          &                     & 8595.4$^{+1.5}_{-1.8}$ &  8.9$^{+7.0}_{-3.0}$\\
7 & 6548 [NII]          & 6.9$^{+2.5}_{-1.8}$ & 8550.1$^{+3.9}_{-3.0}$ &  1.7$^{+1.2}_{-1.0}$\\
  & H$\alpha$ narrow &                     & 8568.5$^{+2.2}_{-1.5}$ &  3.8$^{+1.8}_{-1.0}$\\
  & H$\alpha$ broad  & 144$^{+26}_{-19}$   & 8573.8$^{+13.1}_{-14.2}$ & 23.5$^{+4.2}_{-3.8}$\\ 
  & 6583 [NII]          &                     & 8595.6$^{+1.0}_{-1.0}$ &  6.0$^{+2.1}_{-1.4}$\\
\hline
\end{tabular}
\end{table}

\begin{figure}
\centering
\epsfig{file=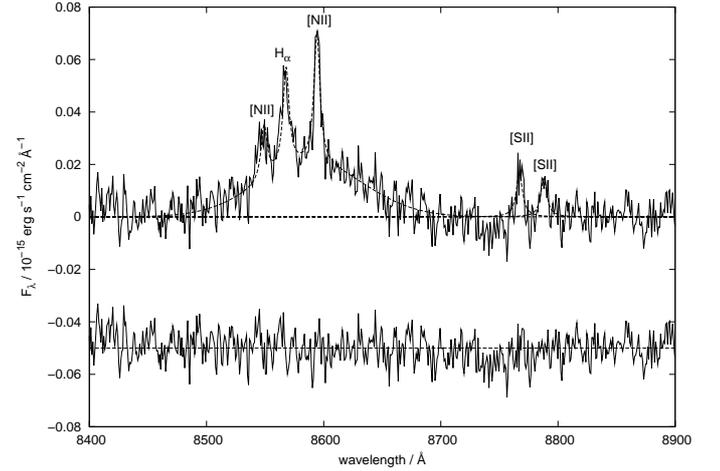,width=9cm}
\caption{\label{viivasovitus} Example of the line fitting (epoch four).
The data are shown as a continuous line and the model fit as a dashed line.
The lower part shows the residuals of the fit, shifted downwards by 0.05
$\times 10^{-15}$.
}
\end{figure}

To study the properties and significance of the emission lines further,
model fits were made to the observed spectra. The narrow lines were
modeled by a Lorentzian profile and the broad H$\alpha$ line with a
Gaussian profile. Only significant features were fitted and the broad
H$\alpha$ was fitted simultaneously with the narrow lines on top of
it. Figure \ref{viivasovitus} shows the result of the line fitting on
epoch four. Table \ref{voima} lists the values of the fitted parameters and
their associated errors.

The error bars were estimated by performing model fits to 500
simulated spectra for each epoch. The simulated data sets contained
all significant emission features for that particular epoch. The noise
in the spectra was simulated by performing a Fourier transform to the
continuum-subtracted pure continuum parts of the observed spectra,
randomizing the phases and transforming back to obtain the simulated
noise. In this way the statistical properties of the noise were
preserved and the fluctuations the simulated spectra resembled closely
the observed fluctuations. Examining the distributions of the
simulation results it was found that for some parameters (e.g. the
FWHM of the narrow lines) the distribution was not symmetrical, but
skewed towards higher values.  For this reason asymmetric error bars
are given in Table \ref{voima}. The quoted error range includes 67\%
of the simulated values, computed separately above and below the
median value.  The broad H$\alpha$ was considered significant only if
the flux of the line could be determined to better than 3$\sigma$
accuracy, where $\sigma$ is the lower error bar. Table \ref{voima}
lists only significant detections of the broad H$\alpha$ (epochs four
and seven). We also computed the 3$\sigma$ upper limits for the broad
H$\alpha$ line by creating simulated data sets of $\sim$ 300
simulations, each containing increasing contribution of the broad
H$\alpha$ and determining the line flux which could be detected with
3$\sigma$ significance.

The rms scatter in the position of the $\lambda$6583 [NII] line is 0.8
\AA, consistent with the $\sim$ 1 \AA\ error bars derived by the error
simulations.  The average redshift derived from the $\lambda$6583
[NII] line is $0.3056 \pm 0.0001$, consistent with the earlier
redshift z = 0.306
\citep{1978bllo.conf..176M,1985PASP...97.1158S}. The error in the
redshift represents the internal precision of the measurements without
systematic effects, which are expected to be small because the
wavelength calibration was derived from sky lines exposed
simultaneously with the OJ~287 spectrum.

\begin{figure}
\centering
\epsfig{file=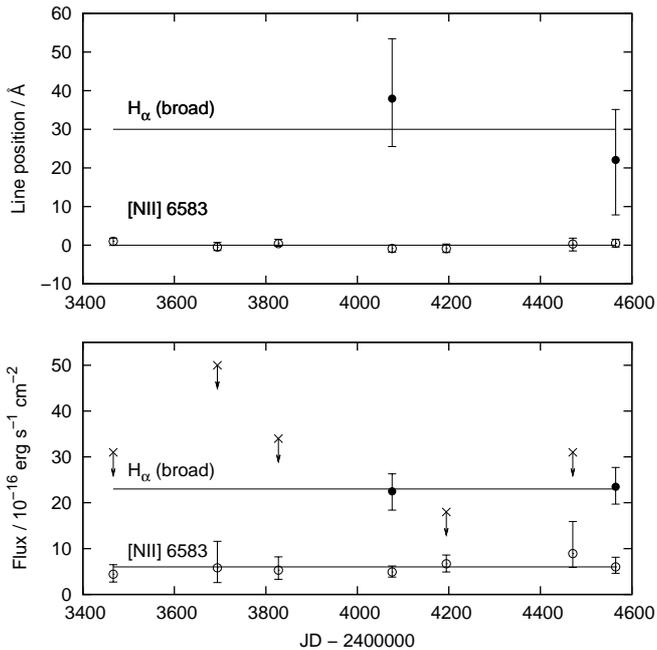,width=9cm}
\caption{\label{viivamuutos} Line positions (upper panel) and flux
density (lower panel) of the broad H$\alpha$ and $\lambda$6583
[NII] lines. In both panels the results for the H$\alpha$ line are
shown with closed symbols (upper limits with crosses) and the
$\lambda$6583 [NII] line with open symbols. Horizontal lines represent
average values.  In the upper panel the the average value has been
subtracted and the plot for the H$\alpha$ moved 30 \AA\ up for
clarity.}
\end{figure}

Figure \ref{viivamuutos} shows the position and flux density of the
$\lambda$6583 [NII] and H$\alpha$ lines. The $\lambda$6583 [NII]
line is not expected to show any variations over the timescales
considered here, which is also confirmed by our results: no
variability larger than the error bars is observed. However, the error
bars of the flux densities are quite large. The broad $H_{\alpha}$
line shows no significant changes in position or flux between the two
epochs it was detected. As the upper limits show, the data are
consistent with no change in the broad H$\alpha$ line flux
throughout the whole monitoring campaign.  Thus we find no evidence of
changes in the BLR in our data.  However, there is a big difference to
the line flux observed by \cite{1985PASP...97.1158S}, who reported a
H$\alpha$ line flux of $2.4 \times 10^{-14}$ erg cm$^{-2}$
s$^{-1}$, which is a factor of $\sim$ 10 higher than in 2005-08. The
difference remains even after subtracting the summed contribution of
the [NII] and narrow H$\alpha$ lines, $12 \times 10^{-16}$ erg
cm$^{-2}$ s$^{-1}$, which were unresolved by
\cite{1985PASP...97.1158S}.

The weighted average of the broad H$\alpha$ line width is $120 \pm
14$ \AA, which corresponds to $4200 \pm 500$ km/s at the redshift of
OJ~287.  \cite{1985PASP...97.1158S} do not give the line width, but
estimating from their Fig. 2 it is $\sim$ 120 \AA, i.e. the same as
observed here. 

\section{Discussion}

\begin{figure}
\centering
\epsfig{file=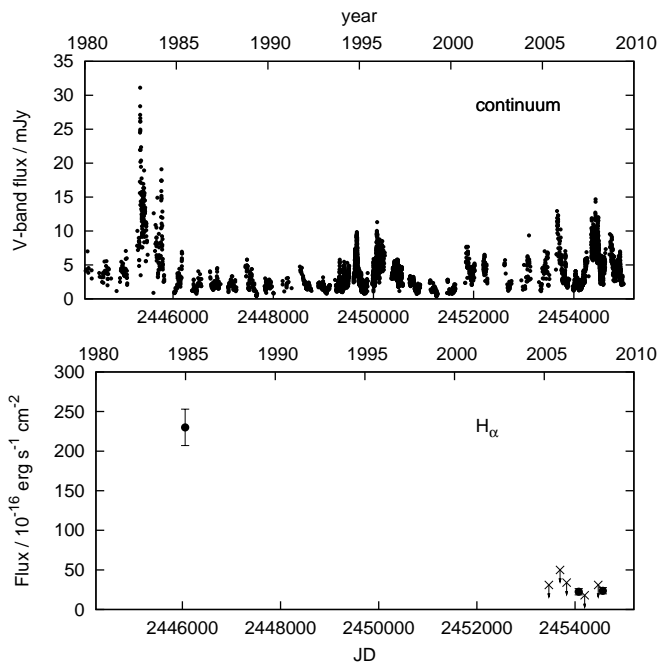,width=9cm}
\caption{\label{kontiviiva} Continuum (upper panel) and
H$\alpha$ (lower panel) flux of OJ~287 in 1980-2010. In the lower
panel crosses with arrows mark 3$\sigma$ upper limits and solid circles
detections. The error bar of the leftmost point in the lower panel
has been arbitrarily set to 10\%.}
\end{figure}

The most notable result of our monitoring is the dramatic decrease in
broad H$\alpha$ luminosity between 1984 and 2005-08. In Fig.
\ref{kontiviiva} we show the continuum and broad H$\alpha$ line flux
of OJ~287 in 1980-2010. In the upper panel three outburst seasons of
OJ~287 can be seen: 1982-84, 1994-95 and 2005-08. The high H$\alpha$
flux appeared just after the strong outbursts in 1982-83, 290 days
after the second peak.  Unfortunately, the sampling is too poor to
draw any conclusions on the possible connection between continuum and
BLR luminosity in OJ~287. Such a connection would not even be
necessarily expected because in BL Lac objects the vast majority of the
continuum arises from the jet and is highly beamed by relativistic
effects due to the small angle between the line of sight and the jet
axis. Thus the observed continuum variations are not necessarily
connected to the changes of the continuum source illuminating the
H$\alpha$ emitting clouds, most likely the inner parts of the
accretion disk. It is nevertheless an intriguing observation that the
high line luminosity in 1984 was observed right after a luminous
continuum outburst. Given the weakness of the broad H$\alpha$ line we
are unable to study a possible connection between BLR velocity field and
the suggested periastron of the secondary black hole in 2005-07
\citep{2007ApJ...659.1074V}.

\begin{figure}
\centering
\epsfig{file=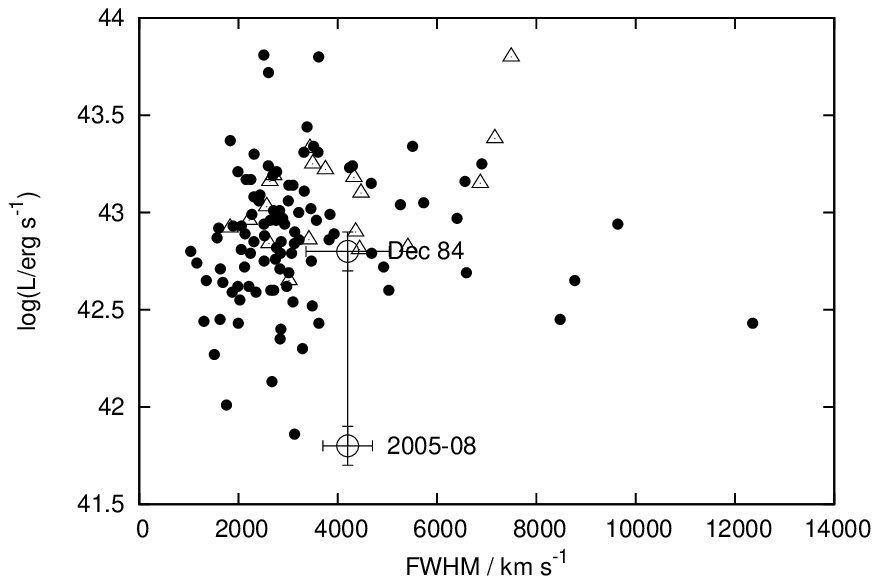,width=9cm}
\caption{\label{lumvertailu} Comparison of broad H$\alpha$
luminosities of OJ~287 (open circles connected by a line), z = 0.30-0.31
quasars from \citet[][filled circles]{2007AJ....134..102S} and
z = 0.36 Seyfert galaxies from \citet[][open triangles]{2008ApJ...673..703M}.}
\end{figure}

The average line luminosities in 2005-08 were $\log(L/{\rm erg\
s}^{-1})$ = 41.8, 41.1, 41.2 and 40.8 for broad H$\alpha$, narrow
H$\alpha$, $\lambda$6583[NII] and $\lambda$6548[NII], respectively.
The H$\alpha$ luminosities of Seyfert galaxies and quasars are
generally much higher as shown by he comparisons in Table
\ref{llumtable} and Fig. \ref{lumvertailu}. The quasar data were
obtained from \cite{2007AJ....134..102S}, who give data of
77~429 quasars from the Fifth Data Release of the Sloan Digital Sky
Survey \citep{2007ApJS..172..634A}. We selected quasars at the
redshift interval z = $0.30 - 0.31$ (147 quasars) and obtained the
H$\alpha$ line fluxes and widths from the SpecLine tables in the
SDSS archive. Figure \ref{lumvertailu} shows only quasars with FWHM $>$
1000 km s$^{-1}$ (101 quasars). In addition we plot data for 17 z =
0.36 Seyfert galaxies in \cite{2008ApJ...673..703M}.  As can be seen
from Fig. \ref{lumvertailu} the H$\alpha$ luminosity of OJ~287 was
lower than in typical quasars and Seyfert galaxies by a factor of
$\sim$ 10 in 2005-08. In December 1984, however, the H$\alpha$
luminosity was comparable to that of quasars and Seyferts.  We finally
note that the luminosities of the two narrow lines $\lambda$6548 and
$\lambda$6583 [NII], however, are comparable to the narrow-line
luminosities of the quasars in \cite{2007AJ....134..102S}.

\begin{table}
\caption{\label{llumtable} Comparison of line luminosities of OJ~287 to
those of quasars and Seyfert galaxies.}
\centering
\begin{tabular}{llll}
\hline
     & \multicolumn{3}{c}{Log L}\\
Line & OJ~287 & Quasars & Seyferts\\
\hline
H$\alpha$ broad & 41.8 & 42.9 $\pm$ 0.4 & 43.1 $\pm$ 0.3\\
$\lambda$6548[NII] & 40.8 & 41.3 $\pm$ 0.8 & \ldots \\
$\lambda$6583[NII] & 41.2 & 41.4 $\pm$ 0.7 & \ldots \\
\hline
\end{tabular}
\end{table}

\section{Summary}

We have presented high S/N spectra of the BL Lac object OJ~287 during
seven epochs in 2005-08. Our results can be summarized as follows:

1) We were able to detect five narrow emission lines,
$\lambda\lambda$6548,6583[NII], $\lambda6563$H$\alpha$ and
$\lambda\lambda$6716,6731 [SII] during at least one of the epochs and
a broad H$\alpha$ feature during two epochs. The luminosities of the
[NII] lines are comparable to those in quasars at the same redshift,
whereas the broad H$\alpha$ line is a factor of $\sim$ 10 less
luminous than the H$\alpha$ line in quasars and Seyfert galaxies.

2) The luminosity of the broad H$\alpha$ line was a factor of $\sim$
10 lower in 2005-08 than in 1984 when it was reported to have been
detected the last time.

3) We do not see any significant change in luminosity, position
or width if the broad H$\alpha$ line between the two epochs it was
detected in 2005-08. The average width is 4200 km s$^{-1}$, consistent
with the width observed in 1984. Due to the weakness of the broad
emission lines we are unable to study the connection between continuum
and line luminosity or between the BLR velocity field and the proposed
periastron of the secondary black hole in 2005-07.

%\begin{acknowledgements}

%\end{acknowledgements}

\bibliographystyle{aa}

\bibliography{14198}

\end{document}